# High-power frequency comb source tunable from 2.7 to 4.2 μm based on difference frequency generation pumped by an Yb-doped fiber laser

GRZEGORZ SOBOŃ,[1,2,*] TADEUSZ MARTYNKIEN,[3] PAWEŁ MERGO,[4] LUCILE RUTKOWSKI,[1] ALEKSANDRA FOLTYNOWICZ[1,5]

[1]*Department of Physics, Umeå University, 901 87 Umeå, Sweden*
[2]*Laser & Fiber Electronics Group, Faculty of Electronics, Wrocław University of Science and Technology, 50-370 Wrocław, Poland*
[3]*Faculty of Fundamental Problems of Technology, Wrocław University of Science and Technology, 50-370 Wrocław, Poland*
[4]*Laboratory of Optical Fiber Technology, Maria Curie-Sklodowska University, pl. M. Curie-Sklodowskiej 3, 20-031 Lublin, Poland*
[5]*E-mail address: aleksandra.foltynowicz@umu.se*
*Corresponding author: grzegorz.sobon@pwr.edu.pl*

**We demonstrate a broadband mid-infrared (MIR) frequency comb source based on difference frequency generation (DFG) in periodically poled lithium niobate (PPLN) crystal. Mid-infrared radiation is obtained via mixing of the output of a 125 MHz repetition rate Yb-doped fiber laser with Raman-shifted solitons generated from the same source in a highly nonlinear fiber. The resulting idler is tunable in the range of 2.7 – 4.2 μm with average output power reaching 237 mW, and pulses as short as 115 fs. The coherence of the MIR comb is confirmed by spectral interferometry and heterodyne beat measurements. Applicability of the developed DFG source for laser spectroscopy is demonstrated by measuring absorption spectrum of acetylene at 3.0 – 3.1 μm.**

Intense development of optical frequency comb sources in the mid-infrared is mainly driven by their applications in laser spectroscopy [1]. The use of MIR combs enables high-speed and accurate detection of various molecular species, since they possess their fingerprints in this wavelength range [2]. For sensitive detection of multiple species at a time, high-power, broadband, and widely tunable comb sources are required. Usually, for this purpose optical parametric oscillators (OPOs) are used, since they provide high output power with broad spectral coverage [3,4]. However, the requirement of a phase lock of the cavity to the pumping laser increases the complexity of the OPO sources, and limits their practical applicability. Mid-infrared comb sources based on difference frequency generation (DFG) are interesting alternatives to OPOs, because of their simplicity, single-pass configuration, broad tunability, and fully passive cancellation of the carrier-envelope offset frequency ($f_{CEO}$) [5]. While the lack of $f_{CEO}$ is a drawback in detection schemes using tight lock of the comb to an enhancement cavity [6], it is advantageous in other cavity-enhanced techniques, like continuous-filtering Vernier spectroscopy [7]. The variety of nonlinear crystals available today enable generation of radiation up to 13-17 μm using GaSe [8], AgGaSe$_2$ [9], CdSiP$_2$ [10], or OP-GaAs [11] with output power at the level of few to tens of mW. However, DFG comb sources based on PPLN crystals are still very attractive because of high conversion efficiency and wide availability of the crystals, even though the transmission window of lithium niobate is limited to 5 μm. For example, mixing of 1 μm with 1.55 μm radiation, easily achievable from state-of-the-art rare-earth-doped fiber lasers, in PPLN crystals allows reaching the important region around 3.3 μm, where the C-H stretch transitions are located. Such DFG frequency comb can be generated either by mixing the output of two independent sources (e.g. a mode-locked oscillator and continuous-wave (CW) laser [12,13]), or - preferably - using one master source. In the latter case, the DFG system can be driven by e.g. an Er-doped fiber laser [14-16]. The pump radiation for the DFG process is then provided by down-conversion of the 1.56 μm signal to the 1 μm band in a highly nonlinear fiber (HNLF) through the so-called dispersive wave (DW) generation (also referred to as Cherenkov radiation [17]). This approach benefits from the high coherence of the DW, however, the power stored in the DW is usually low [18] and an additional Yb-doped fiber amplifier in the pump branch is required to boost the power prior to frequency mixing in the crystal [15,16]. If the amplifier is lacking, the resulting MIR idler power is limited to single miliwatts [14]. A more efficient approach is based on high-power Yb-doped fiber laser that simultaneously serves as pump for DFG, and provides tunable signal via Raman-induced soliton self-frequency shift (SSFS) in a HNLF [19-21]. This approach has led to generation of MIR radiation tunable from 3.0 to 4.4 μm with maximum power of 125 mW, but with no coherence at wavelengths shorter than 3.2 μm [19]. Recent reports show that broadband radiation in the 2.9 – 3.6 μm wavelength range can be achieved directly from the SSFS in fluoride fibers [22]. However, this approach requires a powerful pump laser at 2.8 μm, and the longest achieved wavelength is 3.6 μm to date. Moreover, the coherence properties of such sources, and thus their usability in optical frequency comb spectroscopy were not investigated.

Here, we report a DFG source based on an Yb-doped fiber laser, delivering a coherent idler tunable from 2.7 to 4.2 μm with average output power reaching 237 mW, FWHM bandwidth of 240 nm, and pulses as short as 115 fs. Thanks to the wide tuning range of the Raman solitons generated in the HNLF, the DFG idler can be tuned down to 2.7 μm, which is the lowest ever reported wavelength achieved from a PPLN-based DFG system [15,16,19-21]. Additionally, the power per comb mode is significantly higher than reported previously, which will translate into sensitivity improvement in comb spectroscopy.

The experimental setup of the DFG source is depicted in Fig. 1. The system is driven by an Yb-doped fiber laser (Orange-HP, Menlo Systems), which delivers sub-150 fs pulses centered at 1.04 μm with maximum power of 2.3 W at 125 MHz repetition rate. The laser output is divided into two branches. One part is coupled into a 90 cm-long piece of microstructured silica HNLF (produced in the Laboratory of Optical Fiber Technology, Maria Curie-Sklodowska University, Lublin, Poland), which provides a tunable signal for the DFG via Raman-induced SSFS. The second part is directed through a delay line, which ensures a temporal overlap between the 1040 nm pump and the frequency-shifted signal pulses. Additionally, a telescope in the pump arm optimizes the spot size for best efficiency of the DFG. The beams are combined on a dichroic mirror and focused by an achromatic lens in a 3 mm thick MgO:PPLN crystal (MOPO1-0.5-3, Covesion Ltd.) with 9 available quasi-phase matching (QPM) periods for MIR idler tuning. The output beam is collimated with a calcium fluoride ($CaF_2$) lens. The MIR idler is separated from the unabsorbed near-infrared beams using a 5 mm-thick Germanium filter.

Figure 2 shows the chromatic dispersion curve of the HNLF, together with an image of the fiber cross section taken with a scanning electron microscope (SEM). The silica core with 3.07 μm diameter is surrounded by eight rings of air holes, with the distance between the holes equal to 2.47 μm, and average diameter of the air holes of 1.84 μm. The dispersion was calculated using Comsol Multiphysics Wave Optics Module taking into account the material dispersion of silica glass.

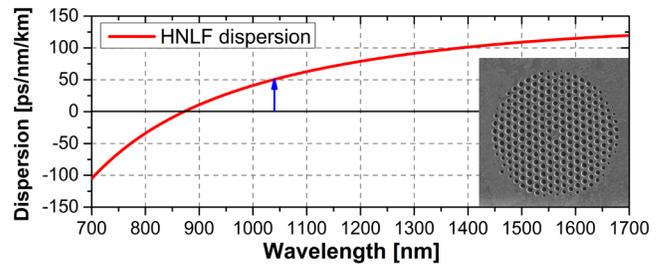

Fig. 2. Calculated dispersion of the HNLF with indicated pump wavelength (blue arrow). Inset: fiber cross-section SEM image.

The signal pulses coming out of the HNLF can be easily shifted up to 1675 nm by increasing the pump power coupled into the fiber. Figure 3 presents examples of signal spectra recorded at different pumping levels. The spectral FWHM exceeds 30 nm in the entire tuning range, reaching 39 nm at 1675 nm central wavelength at the highest pumping power of 500 mW. The output power (stored in the shifted soliton) varies between 35 to 75 mW in the tuning range from 1425 to 1675 nm. During the power measurement, the solitons were filtered out from the spectrum using a 1400 nm long-pass filter. For shifts below 1425 nm the exact power in the pulse could not be measured, because of the second-order soliton appearing at around 1270 nm, and the residual pump.

It has been shown that pulses generated via SSFS are usually chirp-free, with nearly transform-limited duration [23]. Nevertheless, we verified the pulse shape and duration via interferometric autocorrelation using a standard Michelson interferometer with a 2-mm thick BBO crystal for second harmonic generation (SHG). The inset in Fig. 3 shows the autocorrelation trace of a pulse centered at 1500 nm, with duration of 106 fs (assuming a Gaussian pulse shape with deconvolution factor of 0.707). The spectral FWHM at this particular wavelength is 34 nm, which corresponds to a time-bandwidth product (TBP) of 0.48, close to the transform limit (0.441 for a Gaussian pulse).

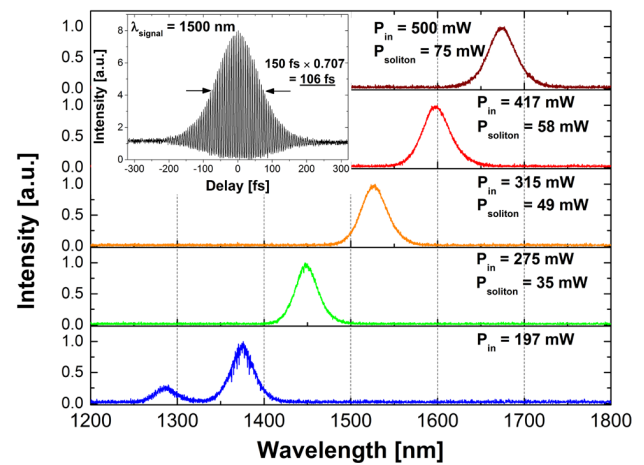

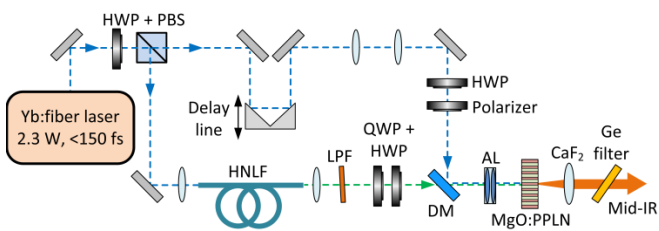

Fig. 1. Setup of the DFG system. HWP: half-wave plate; PBS: polarization beam splitter; QWP: quarter-wave plate; LPF: long-pass filter; DM: dichroic mirror; AL: achromatic lens.

Fig. 3. Spectra of the frequency-shifted solitons for different pumping levels ($P_{in}$) with indicated power stored in the pulse ($P_{soliton}$). Inset: autocorrelation of the pulse centered at 1500 nm (106 fs).

In order to verify the phase coherence of the generated signal pulses we measured the interference between two consecutive pulses in an unequal-path Michelson interferometer, similar to that presented in [24]. This technique is commonly used to verify the coherence of supercontinuum (SC) sources and mode-locked oscillators, since the visibility of the modulation in the interference signal directly corresponds to the degree of coherence of the generated pulses [24,25]. The measured spectral interference patterns at different wavelengths are plotted in Fig. 4. The fringe visibility, defined as $V(\lambda) = [I_{max}(\lambda)-I_{min}(\lambda)]/[I_{max}(\lambda)+I_{min}(\lambda)]$, where $I_{max}$ and $I_{min}$ are the maximum and minimum intensities in the signal, respectively [24-26], is at the level of 0.8-0.9 for shifts up to 1600 nm, which proves that the generated signal pulses are highly coherent. We observe slight degradation of the coherence at longer wavelengths, nevertheless, the visibility at 1650 nm is at the level of >0.65.

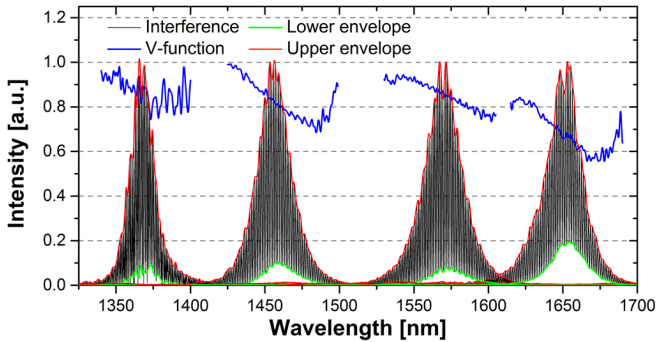

Fig. 4. Spectral interference patterns of consecutive pulses, measured at different wavelengths of the shifted soliton. Blue curve represents the calculated fringe visibility function. Red and green curves show the upper and lower envelopes of the interference patterns, used for obtaining $I_{max}$ and $I_{min}$, respectively.

Representative spectra of the generated MIR idler recorded with a Fourier-transform spectrometer (FTS) are depicted in Fig. 5, together with the average output power measured after the $CaF_2$ lens and the Ge filter (right scale). The tuning range in the DFG process is determined by the phase-matching bandwidth of the crystal and by the available tuning range of the signal. In our experiment, the solitons could be shifted up to 1675 nm, which results in an idler centered at around 2700 nm. At the long wavelength side, the achievable idler wavelength was limited by the periods available in the crystal (shortest period of 27.91 μm provides QPM up to 4.2 μm). The FWHM of the tunable idler varies between 140 to 240 nm. The maximum achievable power is 237 mW at 3.3 μm, obtained with 1.46 W of pump and 50 mW of signal power. This corresponds to around 16% conversion efficiency with respect to the pump. The calculated peak power spectral density is ~1.1 mW/nm, approx. 60% higher than obtained from a similar DFG source reported previously [19]. The average power per comb mode, taking into account the full spectral bandwidth, is ~2.3 μW. It is worth highlighting that high output power >160 mW is maintained over the entire available tuning range. Moreover, we observed no roll-over in the output vs. pump power characteristic, which indicates that the output power can be further scaled.

The temporal properties of the generated MIR pulses were verified by measuring the interferometric autocorrelation using a 1-mm thick PPLN crystal for SHG. The measurements revealed that pulse duration of <200 fs is maintained over the entire tuning range. The shortest pulse of 115 fs (assuming a Gaussian pulse shape) was observed at a wavelength of 3500 nm. The corresponding autocorrelation trace is depicted in Fig. 6. With 210 nm of FWHM bandwidth, the TBP is equal to 0.59, which means that the generated pulses can be further compressed to the transform limit of 85 fs.

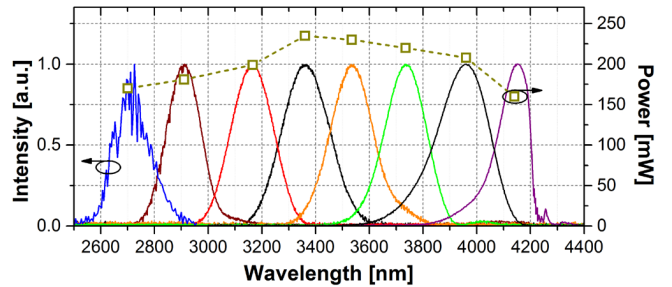

Fig. 5. Normalized MIR idler spectra throughout the full tuning range (left scale), together with the average output power (right scale). The structure in the spectrum centered at 2700 nm is caused by strong water absorption lines present in this spectral range.

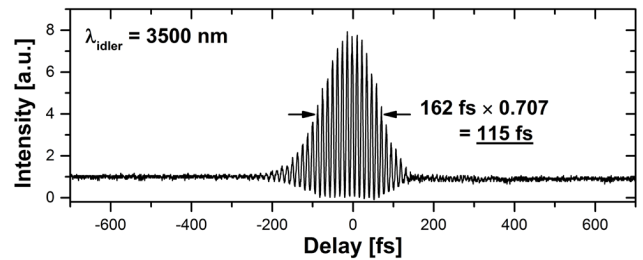

Fig. 6. Interferometric autocorrelation of the idler pulse centered at 3500 nm, revealing pulse duration of 115 fs.

The coherence of the generated idler was additionally examined by performing heterodyne beat between the comb and a reference CW laser. This technique is also commonly used as a coherence test of SC [27] and DFG sources [15]. Since no MIR CW laser was available, we frequency-doubled the MIR idler centered at around 3100 nm using the 1-mm thick PPLN, resulting in ~3 mW of SHG radiation. The beat between the free-running frequency-doubled idler comb and a 1577 nm Er-doped fiber laser with <1 kHz linewidth and power of 25 mW was detected using an InGaAs photodiode. Figure 7(a) presents the beat notes measured with 100 kHz resolution bandwidth. For comparison, we also measured the beat between the CW laser and the signal comb (taken directly at the output of the HNLF), which is depicted in Fig. 7(b). The corresponding beat notes are stronger than those in Fig. 7(a) because of the much higher power available in the NIR signal (average power of the soliton centered at 1577 nm was at the level of 52 mW, compared to 3 mW in the frequency-doubled MIR comb). The 3 dB linewidths of both beat notes are narrower than 400 kHz and are mostly affected by the fluctuations of the repetition rate of the free-running Yb-doped fiber pump laser. Nevertheless, the FWHM of the beat is more than 10 times narrower than reported previously for a similar source [15].

Finally, to demonstrate the suitability of the DFG source for spectroscopic applications, we measured an absorption spectrum of the $v_3$ and $v_2 + (v_4 + v_5)^0_+$ bands of 800 ppm of $C_2H_2$ in $N_2$ at atmospheric pressure, contained in a 23.5 cm-long cell, using an FTS with a resolution of 1.5 GHz. Figure 8 shows the measured absorption spectrum (black) together with the model spectra of 800 ppm of $C_2H_2$ (red) and 0.33% of water present in the 210 cm-long free space path (blue, both inverted), calculated using line parameters from the HITRAN database [28]. The agreement between the measurement and simulation is good. The relatively high noise on the baseline (for 7 minute averaging time) is caused by the lack of active stabilization of the delay line in the pump arm, thus much better performance is expected once it is stabilized.

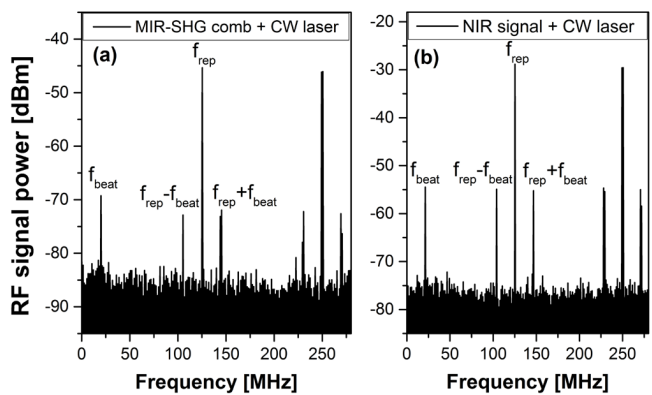

Fig. 7. Beat notes between the narrow-linewidth 1577 nm CW laser and: (a) the NIR signal comb, (b): the frequency-doubled MIR comb.

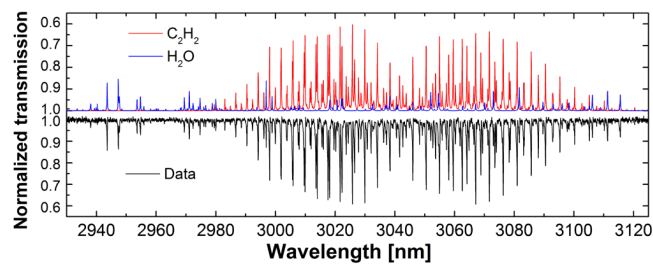

Fig. 8. Absorption spectrum of 800 ppm of acetylene in a cell and 3300 ppm of atmospheric water (black) in the free space path, compared to the model spectra (acetylene – red, water – blue, inverted for clarity).

Summarizing, we have demonstrated a high-power mid-infrared frequency comb source delivering sub-200 fs pulses in the wavelength tuning range of 2.7 – 4.2 μm at 125 MHz repetition rate. The maximum achieved idler power is 237 mW, which is the highest power ever reported from a PPLN-based DFG source in this spectral range. The coherence of the source was verified in the entire tuning range by measuring the fringe visibility of the interference signal of two consecutive signal pulses, as well as by performing a heterodyne beat between the MIR comb (set at central wavelength of 3100 nm) and a reference CW laser. The high average output power (comparable to that obtained from OPO systems, but in a much simpler layout) combined with a wide tuning range, large FWHM and smooth envelope, make this a powerful source for broadband spectroscopy. In the first demonstration, we measured a broadband spectrum of acetylene at 3.0 – 3.1 μm with an FTS. In our future work we will combine this DFG source with a continuous-filtering Vernier spectrometer [29], which will allow fast and robust detection of multiple atmospheric species with concentration detection limits below ppb.

**Funding.** Knut and Alice Wallenberg Foundation (KAW 2015.0159); Faculty of Science and Technology, Umeå University.

**Acknowledgment**. We thank Marco Marangoni and Axel Ruehl for fruitful discussions on the design of the DFG system. We thank Gang Zhao and Thomas Hausmaninger for providing the Er-doped fiber CW laser for heterodyne beat measurement.